\documentclass[runningheads]{llncs}
\usepackage{graphicx}
\usepackage{epstopdf}
\usepackage[colorlinks=true,linkcolor=black,anchorcolor=black,citecolor=black,filecolor=black,menucolor=black,runcolor=black,urlcolor=black]{hyperref}
\begin{document}
	\title{Augmenting a Physics-Informed Neural Network for the 2D Burgers equation by addition of solution data points}
\titlerunning{Augmenting a PINN by addition of solution data points}
%
\author{Marlon S. Mathias\inst{1,2}\orcidID{0000-0002-2415-5723} \and
	Wesley P. de Almeida\inst{1,3}\orcidID{0000-0002-3253-0191} \and
	Marcel R. de Barros\inst{1,3}\orcidID{0000-0002-0759-2497} \and
	Jefferson F. Coelho\inst{1,3}\orcidID{0000-0002-5421-0789} \and
	Lucas P. de Freitas\inst{1,3}\orcidID{0000-0002-3440-1633} \and
	Felipe M. Moreno\inst{1,3}\orcidID{0000-0001-9611-1116} \and
	Caio F. D. Netto\inst{1,3}\orcidID{0000-0002-6627-4931} \and
	Fabio G. Cozman\inst{1,3}\orcidID{0000-0003-4077-4935} \and
	Anna H. R. Costa\inst{1,3}\orcidID{0000-0001-7309-4528} \and
	Eduardo A. Tannuri\inst{1,3}\orcidID{0000-0001-7040-413X} \and
	Edson S. Gomi\inst{1,3}\orcidID{0000-0003-1267-9519} \and
	Marcelo Dottori\inst{1,4}\orcidID{0000-0003-2382-4136}}
\authorrunning{Mathias et al.}
%
\institute{Center for Artificial Intelligence (C4AI) -- University of Sao Paulo, Brazil\\
	\email{\{marlon.mathias,wesleyalmeida,marcel.barros,jfialho,lfreitasp2001,\\felipe.marino.moreno,caio.netto,fgcozman,anna.reali,eduat,gomi,mdottori\}@usp.br}\\
	Av. Prof. Lúcio Martins Rodrigues, 370, Butantã, São Paulo, CEP 05508-020\\
	\url{https://c4ai.inova.usp.br/}\and
	Instituto de Estudos Avançados -- University of São Paulo, Brazil \and
	Escola Politécnica -- University of Sao Paulo, Brazil \and
	Instituto Oceanográfico -- University of Sao Paulo, Brazil}
\maketitle              
\begin{abstract}
	We implement a Physics-Informed Neural Network (PINN) for solving the two-dimensional Burgers equations. This type of model can be trained with no previous knowledge of the solution; instead, it relies on evaluating the governing equations of the system in points of the physical domain. It is also possible to use points with a known solution during training. In this paper, we compare PINNs trained with different amounts of governing equation evaluation points and known solution points. Comparing models that were trained purely with known solution points to those that have also used the governing equations, we observe an improvement in the overall observance of the underlying physics in the latter. We also investigate how changing the number of each type of point affects the resulting models differently. Finally, we argue that the addition of the governing equations during training may provide a way to improve the overall performance of the model without relying on additional data, which is especially important for situations where the number of known solution points is limited.
	
\end{abstract}

\section{Introduction}

The rapid development of Machine Learning (ML) models has led to significant progress in various areas and allowed a new approach to problems involving time-series forecasts. By training these models with large datasets, the models can recognize complex patterns and address the prediction of physical systems. However, a purely data-driven approach may also not be entirely desirable, as such models require substantial amounts of data to train, which might not always be available, and they may lead to non-physical solutions when presented to previously unseen scenarios \cite{karniadakis2021}.


Physics-Informed Machine Learning models present a compromise between purely data-driven and purely physics-based approaches in an attempt to combine their advantages and minimize their shortcomings. The literature presents several ways of combining both models \cite{kashinath2021,raissi2019,willard2020}. These approaches include, but are not limited to: using ML to estimate the error in the physics-based models \cite{xu2015}; using ML to increase the resolution of known flow fields \cite{nair2019,fukami2020}; substituting the governing equations by trained neural networks \cite{wu2020}; adding physical constraints to the ML model \cite{beucler2021,read2019}.

In this work, we implement a Physics-Informed Neural Network (PINN) that uses the governing equations of the physical system in its fitness evaluation. \cite{li2021} calls this approach a neural-FEM, comparing it to the Finite Elements Method (FEM) of solving partial differential equations. In this analogy, the neural network works as one large and complex finite element, which spans the whole domain and solves the equation within its solution space, similarly to the classical FEM, in which several elements, each with a relatively simple solution space, are spread along with the domain in a mesh.

In this paper, we will find solutions for the Burgers equation in a two-dimensional space. Computational Fluid Dynamics researchers widely use this equation as a toy problem intended to test novel ideas that may be used to improve fluid dynamics solvers.

One advantage of PINN models over regular physics-based models is that the neural network may be trained using the governing equations and a set of points where the solution is known beforehand, comprising a hybrid solution between physics-driven and data-driven models. This paper compares neural networks that were trained solely by the governing equations, data points, or an array of combinations between both scenarios.

In real-world uses of machine learning models, there are situations where a limited amount of data is available for training. By adding physical knowledge to the model, it might be possible to reduce the amount of data needed without impacting its quality. Furthermore, it may aid the model in making accurate predictions even for previously unseen situations.

The remainder of this paper is structured as follows: In Section~\ref{sec:methods}, we present the governing equations of our physical system and its boundary conditions and describe the PINN implementation and training procedure. Section~\ref{sec:results} shows the results obtained after training was complete. Finally, Section~\ref{sec:discussion} finishes with some conclusions and suggestions for future works.

\section{Methods}
\label{sec:methods}

We begin this section by describing the governing equations of the physical system and its initial and boundary conditions. Then, we proceed to define the neural network and the loss function that is minimized during training.

\subsection{Governing Equations}

The Burgers equation with two spatial dimensions is given by the system:

\begin{equation}
	\frac{\partial U}{\partial t} + U \frac{\partial U}{\partial x} + V \frac{\partial U}{\partial y} = \nu \left(\frac{\partial^2 U}{\partial x^2} + \frac{\partial^2 U}{\partial y^2} \right) ,
	\label{eq:gov1}
\end{equation}

\begin{equation}
	\frac{\partial V}{\partial t} + U \frac{\partial V}{\partial x} + V \frac{\partial V}{\partial y} = \nu \left(\frac{\partial^2 V}{\partial x^2} + \frac{\partial^2 V}{\partial y^2} \right) ,
	\label{eq:gov2}
\end{equation}

\noindent where $x$ and $y$ are the spatial coordinates, and $U$ and $V$ are the velocities in each direction, respectively. $t$ is the solution time, and $\nu$ is the viscosity. If $\nu$ is set to zero, the solution would eventually lead to extremely high gradients, akin to shock waves in compressible fluids, by setting a positive value to $\nu$, the right-hand side of both equations allows some dissipation, which leads to a smoother solution field.

The solution will be evaluated in the domain $0\le t \le 1$, $0\le x \le 1$, $0\le y \le 1$, with Dirichlet boundary conditions setting both velocities to zero at the domain boundaries. The initial condition is given by:

\begin{equation}
	U(t=0,x,y) = \sin(2\pi x) \sin(2\pi y), 
\end{equation}

\begin{equation}
	V(t=0,x,y) = \sin(\pi x) \sin(\pi y) .
\end{equation}

\begin{figure}
	\includegraphics[width=\textwidth]{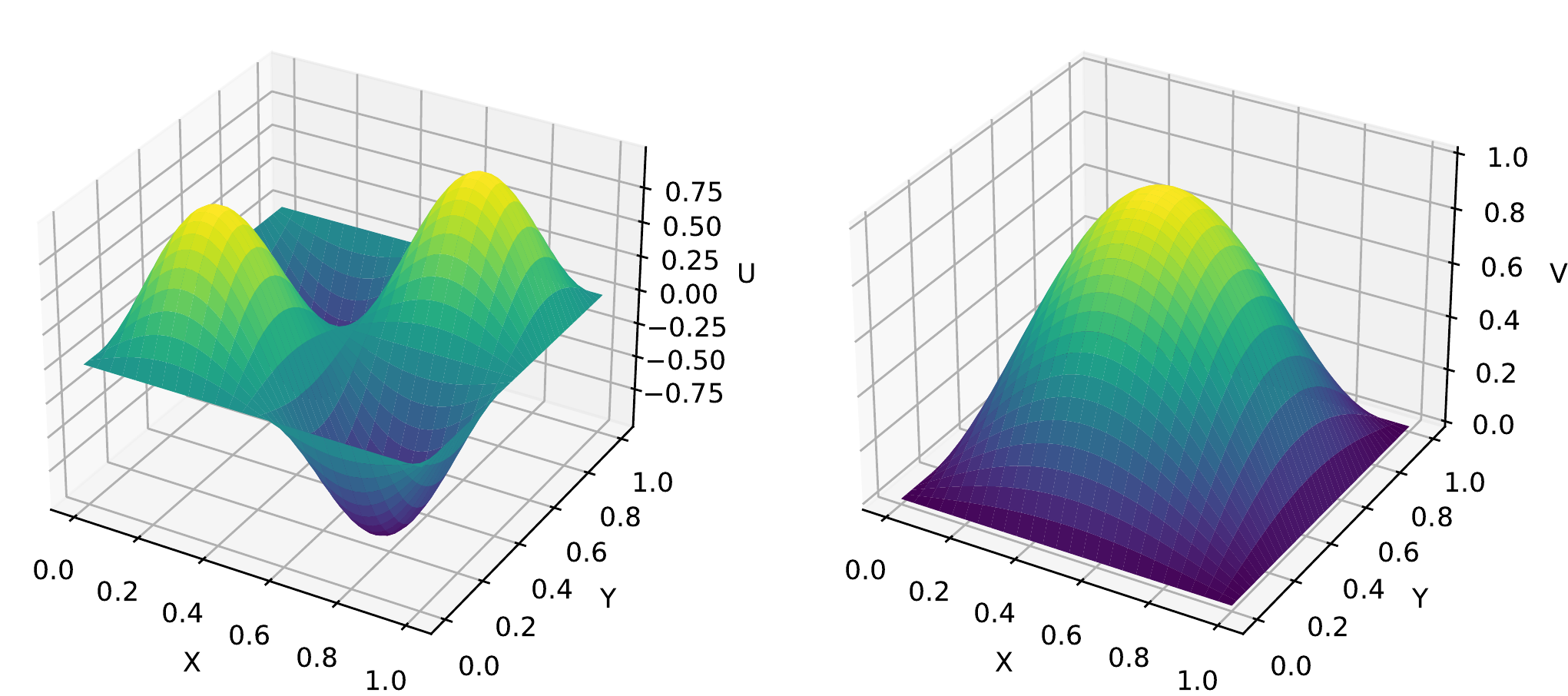}
	\caption{Initial condition for $U$ and $V$.} \label{fig:initial_condition}
\end{figure}

\noindent Figure~\ref{fig:initial_condition} shows $U$ and $V$ at this condition.

\subsection{Physics-Informed Neural Network}

The neural network in our implementation is a Multilayer Perceptron (MLP) and receives coordinates $t$, $x$, and $y$ and outputs the variables $U$ and $V$ for any point of the domain. The MLP network is fully connected and has 4 hidden layers of 20 neurons each, with a hyperbolic tangent activation function. Larger amounts of layers and neurons were tested with little impact on the results. Residual connections are added every two layers, which allows for better computations of the gradients and aids the convergence during training \cite{he2016}. There are two outputs: $\tilde{U}$ and $\tilde{V}$, one for each variable in the problem. The tilde denotes that these variables are the direct output of the MLP and may not observe the boundary conditions of the problem.

After the MLP, there is a Boundary-encoded output layer, as described by \cite{sun2020}. This layer makes sure that the Dirichlet boundary conditions are always observed. The layer works by combining the output of the MLP to a known function that perfectly observes the Dirichlet boundary condition in the following manner: 

\begin{equation}
	\label{eq:networkOutput}
	U = d(t,x,y) \tilde{U} + \left(1-d(t,x,y) \right) U_p ,
\end{equation}

\noindent where $U_p$ is the particular solution for the velocity in the $x$ direction and $d$ is a distance function, which is continuous and differentiable and is equal to zero, where the value of $U$ is set by either a Dirichlet boundary condition or by the initial condition and non-zero everywhere else. The concept behind this definition is that the value of $U$ is fixed to $U_p$ at the boundary and initial conditions but remains under the control of the neural network everywhere else in the domain. This operation is analogous to $V$. We have defined the following equations for $d$:

\begin{equation}
	d = 16 x (1-x) y (1-y) \tanh(\alpha t) .
	\label{eq:distance}
\end{equation}

The hyperbolic tangent function is used so that the value of $d$ is constant and close to one when away from the initial condition. $\alpha$ is set to 26.4, which causes $d$ to reach 0.99 at $t=0.1$. The constant 16 is used, so $d$ becomes close to 1 at the center of the domain. In case other types of boundary conditions, such as Neumann, were present, they would have to be implemented via an additional loss function during training, as they cannot be set by using this method, similarly to the implementation of all boundary conditions by \cite{raissi2019}.

We have set the particular solution of this case as $U$ and $V$ being constant in time and equal to the initial condition values. Therefore:

\begin{equation}
	U_p(t,x,y) = \sin(2\pi x) \sin(2\pi y) .
\end{equation}

\begin{equation}
	V_p(t,x,y) = \sin(\pi x) \sin(\pi y) .
\end{equation}

\begin{figure}[!ht]
	\centering
	\includegraphics[width=1\textwidth]{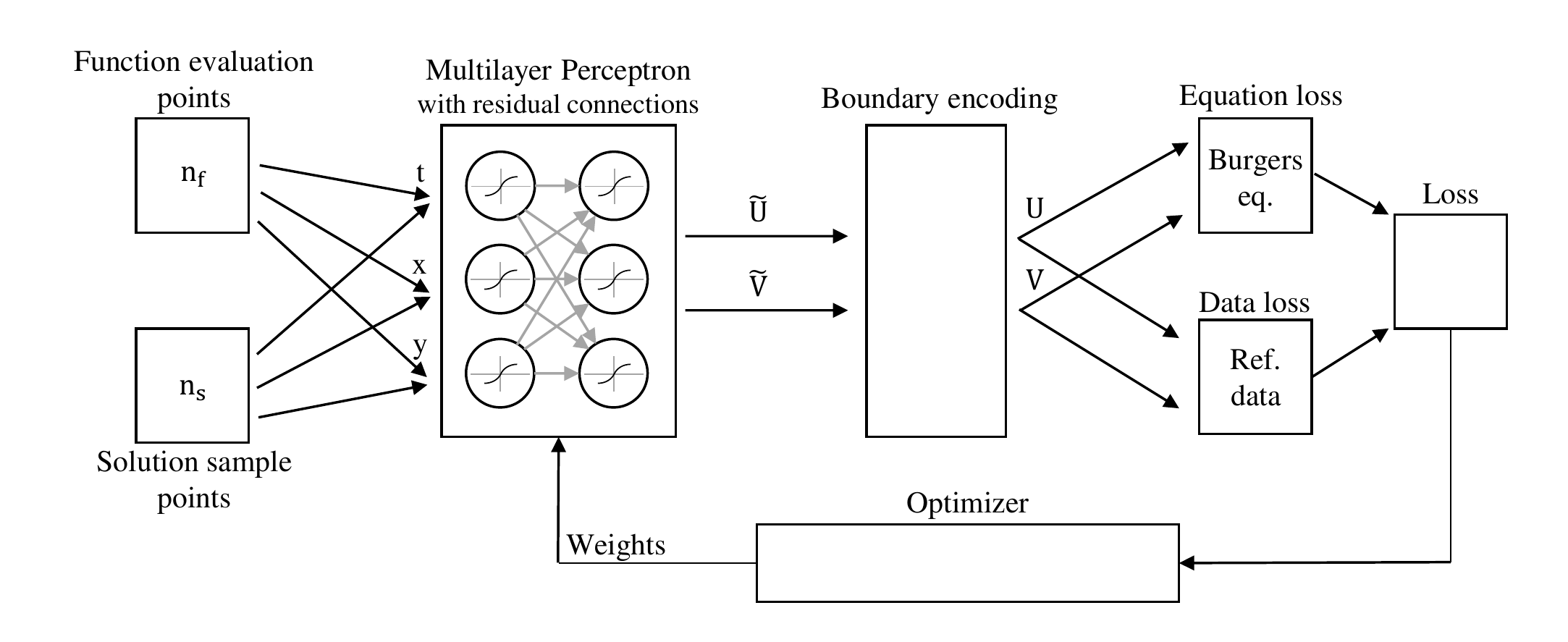}
	\caption{Summary of the PINN network.}
	\label{fig:network}
\end{figure}

In summary, Figure~\ref{fig:network} shows the final configuration of the network.

\subsection{Loss function}

In this work, there are two distinct loss functions whose values are added to obtain the overall loss. First, we will define the loss function relative to the observance of the governing equations, and its value is defined as the sum of the mean absolute value of the residual of each governing equation.

Therefore, given that $U_i$ and $V_i$ are the PINN's output at coordinate ($t_i$,$x_i$,$y_i$), as given by equation~\ref{eq:networkOutput} and its analogous for $V$, the residuals relative to the governing equations are computed as:

\begin{equation}
	R_{f,1,i} = \left| \frac{\partial U_i}{\partial t} + U_i \frac{\partial U_i}{\partial x} + V_i \frac{\partial U_i}{\partial y} - \nu \left(\frac{\partial^2 U_i}{\partial x^2} + \frac{\partial^2 U_i}{\partial y^2} \right) \right|
	\label{eq:r1}
\end{equation}

and

\begin{equation}
	R_{f,2,i} = \left| \frac{\partial V_i}{\partial t} + U_i \frac{\partial V_i}{\partial x} + V_i \frac{\partial V_i}{\partial y} - \nu \left(\frac{\partial^2 V_i}{\partial x^2} + \frac{\partial^2 V_i}{\partial y^2} \right) \right| .
	\label{eq:r2}
\end{equation}

The second type of loss is relative to the data-driven part of the training and is given by the mean absolute distance between the predicted values and known solution values. Naturally, this loss depends on previous knowledge of the solution at given domain points.

The loss function of the model is given by:

\begin{equation}
	\mathcal{L} = \frac{1}{n_f}\sum_{i=1}^{n_f} \left( R_{f,1,i} + R_{f,2,i} \right) + \frac{1}{2n_s}\sum_{i=1}^{n_s} \left( \left| U_i - U^{gt}_i \right| + \left| V_i - V^{gt}_i \right| \right).
\end{equation}

\noindent Where $U^{gt}_i$ and $V^{gt}_i$ are the ground truth values of $U$ and $V$ at point $i$. $n_s$ is the number of points with known solutions. The governing equations are evaluated at $n_f$ random points uniformly distributed along with the domain. One characteristic of PINNs is that there is an infinite pool of points for the loss to be evaluated. Consequently, there is no limit for the value of $n_f$, while $n_s$ is limited to the set of points where the solution is known beforehand.

\subsection{Implementation}

Our code is written in Python and uses the PyTorch module for machine learning tasks. The governing equations' partial derivatives are obtained using PyTorch's autograd feature. The optimization is performed with the Adam algorithm. All tests were executed on a Nvidia\textsuperscript{\textregistered} GeForce\texttrademark~RTX3080 graphics card, with 10 GB of available memory. The code is available on GitHub\footnote{\url{https://github.com/C4AI/PINN\_for\_Burgers}}.

\subsection{Generation of reference data}
\label{sec:met:reference_data}

The data-driven part of the models requires previous knowledge of the solution at some points in the domain. Therefore, we have used numerical differentiation and integration techniques to approximate a solution for the 2D Burgers equation. For this, a mesh of 401 nodes in each direction was uniformly distributed in $x$ and $y$. A sixth-order compact finite differences scheme \cite{lele1992} was used to approximate the spatial derivatives of the governing equations, while a fourth-order Runge-Kutta scheme was used to integrate the solution through time, with a time step of $10^{-5}$. For training, $n_s$ points are randomly chosen from this solution. This model was implemented in Matlab\texttrademark  and will also be available on Github at the time of publication.

\section{Results}
\label{sec:results}

In this section, we compare the solutions obtained by models trained with an array of values for the number of function evaluation points ($n_f$) and the number of solution sample points ($n_s$). We have chosen five values for each variable: 0, 100, 1000, 10000, and 100000. All combinations of values were run with the obvious exception of $n_f=n_s=0$. When $n_f=0$, no knowledge of the governing equation is used, and training is performed solely from the reference data. Similarly, when $n_s=0$, no reference data is used, and the training is purely physics-based.

These cases will also be compared against our reference data, which was generated by more conventional techniques, as described in Section~\ref{sec:met:reference_data}. For all cases, the value to $\nu$ is set to $0.01/\pi$.

\subsection{Sample Results}

\begin{figure}[!ht]
	\centering
	\includegraphics[width=1\textwidth]{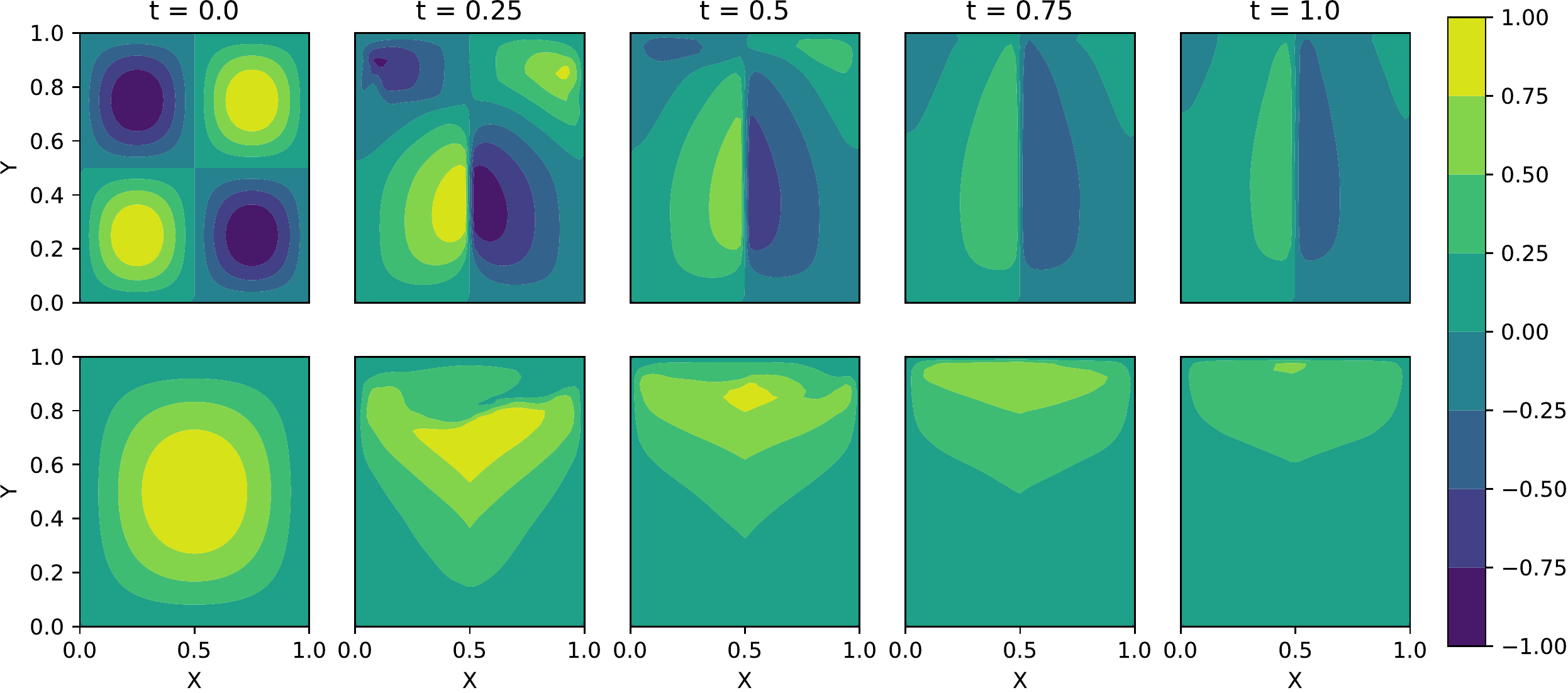}
	\caption{Contours of $U$ (top) and $V$ (bottom) for five equally spaced instants from $t=0$ to $t=1$ (left to right), as modeled by a PINN trained with $n_f=n_s=10^5$.}
	\label{fig:velocity_contours}
\end{figure}

Figure~\ref{fig:velocity_contours} shows contours of both velocity components as modeled by a PINN trained with $10^5$ samples of each type. As the field evolves in time, large gradients form in the middle of the $x$ domain for the $U$ component and at the end of the $y$ domain for the $V$ component. This behavior is expected for the Burgers equation and is representative of shock waves in compressible gasses. The dissipative part, corresponding to the right-hand side of equations \ref{eq:gov1} and \ref{eq:gov2}, causes these large gradients to dissipate as time passes, as can be seen in the images to the right-hand side of the figure. If $\nu$ were set to zero, we would be solving the inviscid Burgers equation, which leads to larger and larger gradients, which are notoriously hard to solve with the techniques presented in this paper.

\subsection{Comparison to Reference Data}

\begin{figure}[!ht]
	\centering
	\includegraphics[width=1\textwidth]{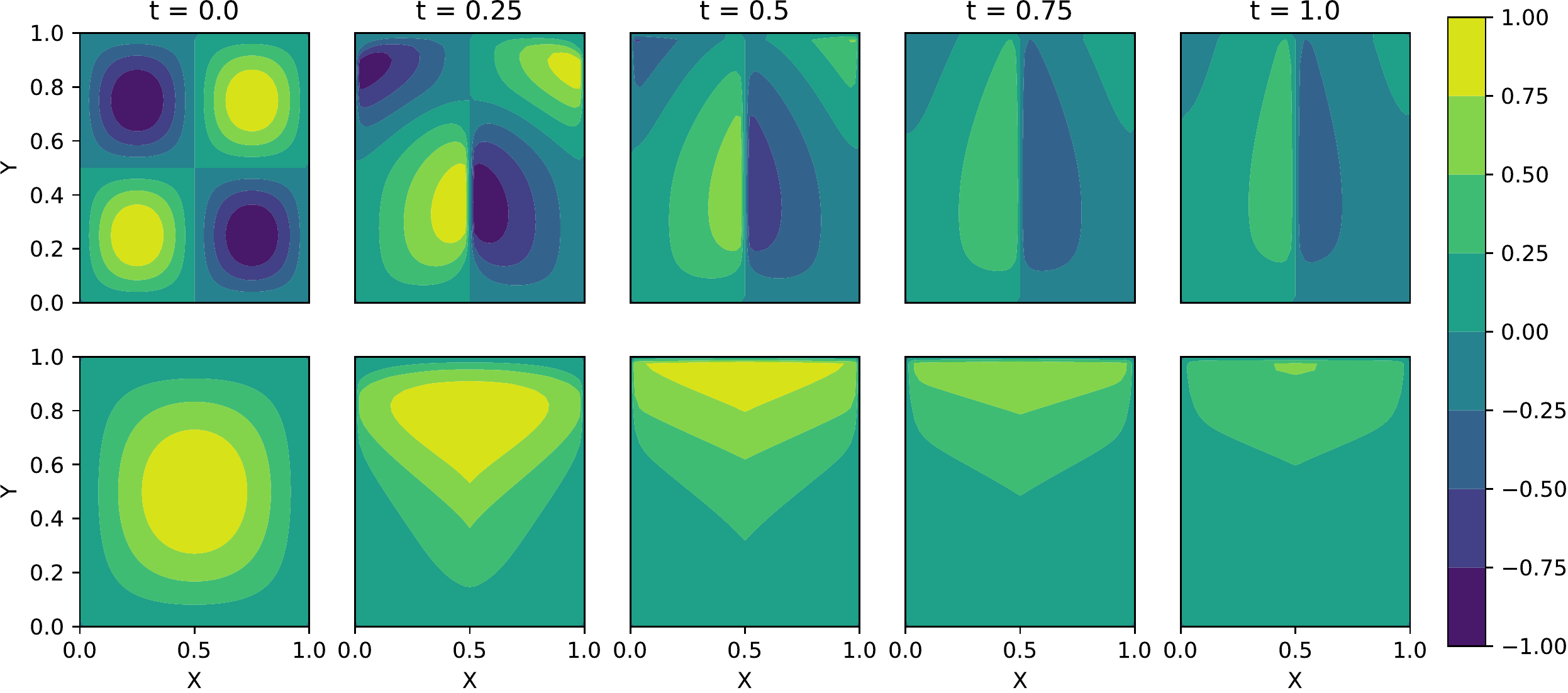}
	\caption{Contours of $U$ (top) and $V$ (bottom) for five equally spaced instants from $t=0$ to $t=1$ (left to right), as modeled by the finite differences numerical solver.}
	\label{fig:velocity_contours_ref}
\end{figure}

The numerical solution described in Section~\ref{sec:met:reference_data} was evaluated in a mesh of 401 by 401 equally spaced nodes, with a time step of $10^{-5}$. Figure~\ref{fig:velocity_contours_ref} shows the velocity contours for the reference data.

\begin{figure}[!ht]
	\centering
	\includegraphics[width=0.8\textwidth]{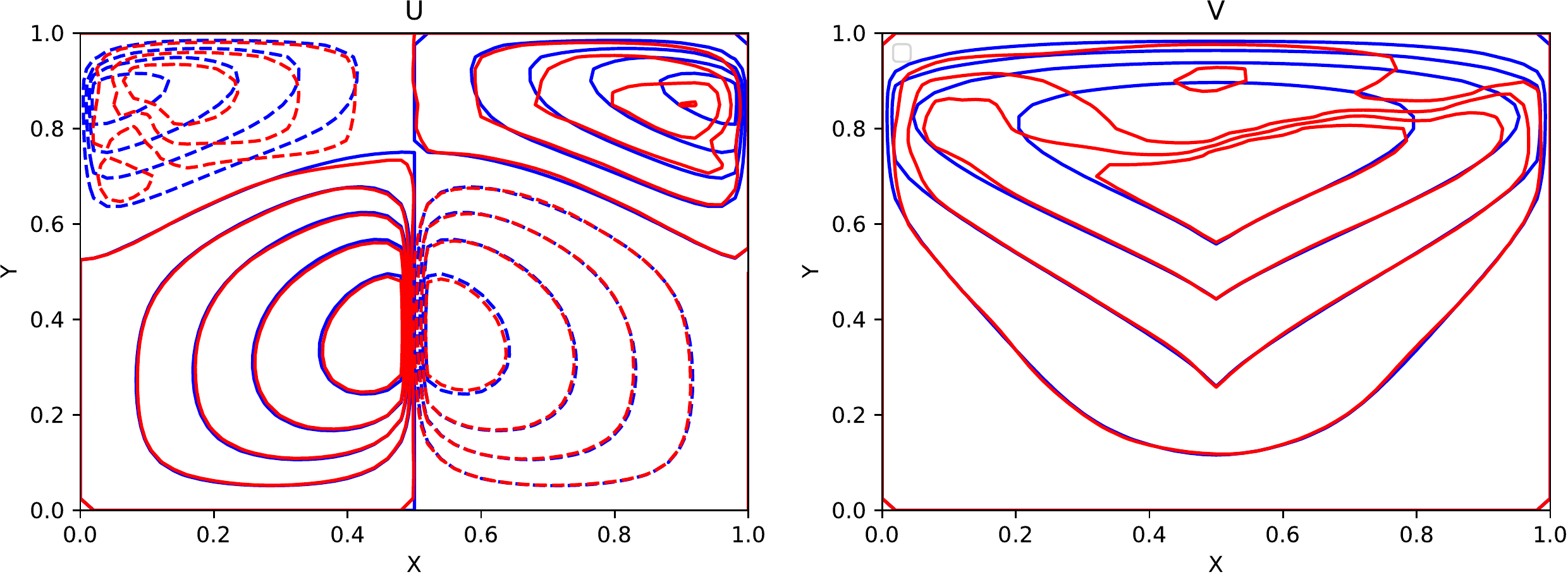}
	\caption{Contours of $U$ (left) and $V$ (right) at $t=0.25$ for the reference data, in blue, and for the PINN with $n_f=n_s=10000$, in red.}
	\label{fig:contour_compare}
\end{figure}

The solution by the PINN and by the finite differences solver -- Figs.~\ref{fig:velocity_contours}~and~\ref{fig:velocity_contours_ref}, respectively -- are visually similar for most of the domain. Nonetheless, both solutions have a clear difference near the $Y=1$ boundary, especially for the $V$ velocity. This variable presents large gradients in this region, physically equivalent to a shock wave in a compressible gas. Interestingly, a similar region of large gradients for $U$ near $X=0$ has reached a much better agreement between the PINN and the reference data. Figure~\ref{fig:contour_compare} overlaps contours of both cases at $t=0.25$. This result leads us to believe that there might be some influence on the implementation of the boundary conditions; perhaps the function chosen for $d(t,x,y)$ in Equation~\ref{eq:distance} is too smooth near the boundaries and makes it difficult for the MLP to model large gradients in this region. Further investigation on the optimal choice of $d(t,x,y)$ is needed.

\subsection{Governing Equations Residuals}

\begin{figure}[!ht]
	\centering
	\includegraphics[width=1\textwidth]{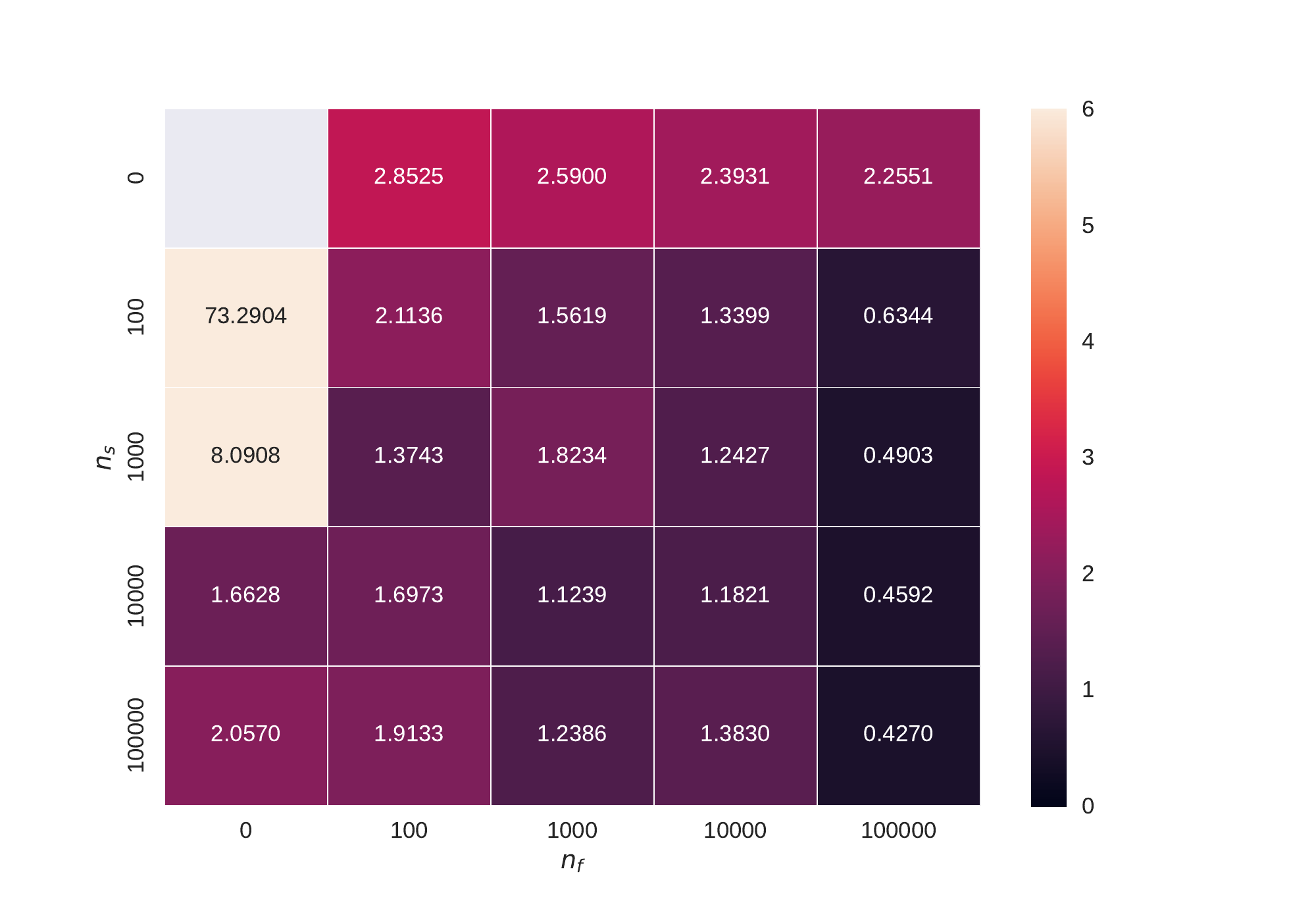}
	\caption{Root mean square of the governing equations' residuals for different combinations of function evaluation points ($n_f$) and solution sample points ($n_s$).}
	\label{fig:heat_residual}
\end{figure}

The PINN was trained for all combinations of $n_f$ and $n_s$ in the range 0, 100, 1000, 10000 and 100000, with the natural exception of $n_f=n_s=0$. After training, the governing equation residuals were measured on a uniformly spaced grid of points across the domain in its three dimensions, one temporal and two spatial. Figure~\ref{fig:heat_residual} shows the root mean square of the residuals for each combination. It is possible to reduce the residual and thus improve the model by increasing either $n_f$ or $n_s$. One significant result we can observe in this figure is when we increase $n_f$ from 0 to 100 at the lower values of $n_s$, such as 100 and 1000. This result illustrates that adding physical knowledge ($n_f > 0$) to cases where few data points are known may be a key to increasing the overall model's accuracy without additional data.

\begin{figure}[!ht]
	\centering
	\includegraphics[width=1\textwidth]{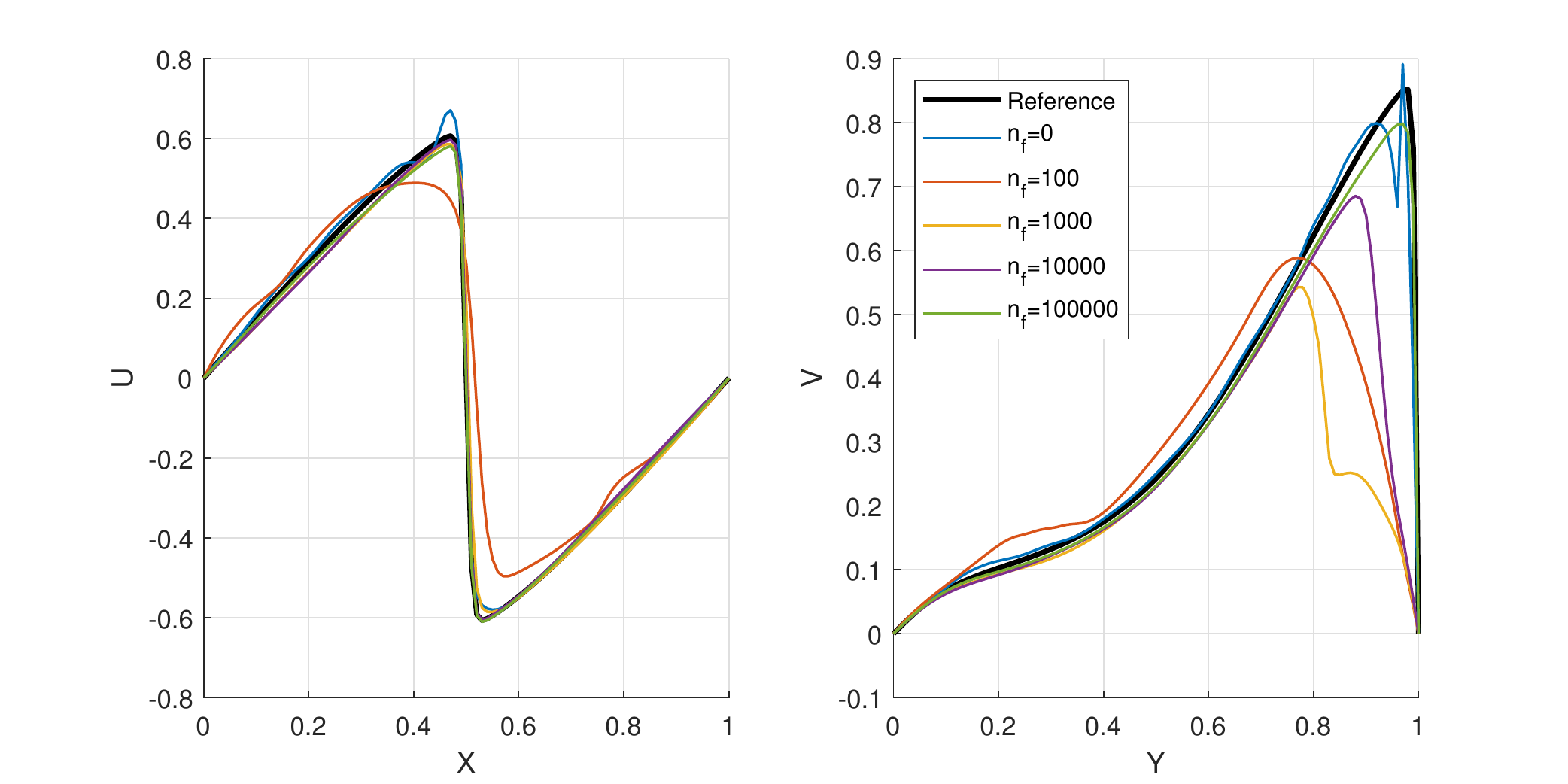}
	\caption{(Left) Plot of $U$ with respect to $x$ at $y=0.25$ and $t=0.5$ for $n_s=1000$ and various values of $n_f$. (Right) Plot of $V$ with respect to $y$ at $x=0.25$ and $t=0.5$ for $n_s=1000$ and various values of $n_f$.}
	\label{fig:nf_compare}
\end{figure}

For further understanding the effect of adding physical knowledge to a data-based approach, we have plotted the value of $U$ along the $x$ axis (with $y=0.25$ and $t=0.5$) for various values of $n_f$, and $n_s$ fixed at 1000. This plot can be seen on the left-hand side of Figure~\ref{fig:nf_compare}. Interestingly, the case with no physical knowledge ($n_f=0$) follows the reference data more closely than some cases with physical knowledge, despite the higher residual. Nonetheless, the higher residual can be easily explained by noting that the values of $U$ for the $n_f=0$ case are not as smooth as those with $n_f>0$. Therefore, despite coming close to the absolute value of the reference data, the derivatives are considerably off, which directly influences the residuals of the governing equations. For $n_f=100$, the curve is much smoother than for the $n_f=0$ case, and, despite being considerably further from the reference data in absolute values, the residuals are lower. For $n_f=1000$ or higher, the curves are much closer to the reference data both in terms of absolute values and in terms of its derivative. The right-hand side of Figure~\ref{fig:nf_compare} shows a similar plot, but of $V$ with respect to $y$ at $x=0.25$ and $t=0.5$. This time, only the $n_f=100000$ case can approach the reference data in terms of absolute values, but the overall conclusion remains.

\begin{figure}[!ht]
	\centering
	\includegraphics[width=1\textwidth]{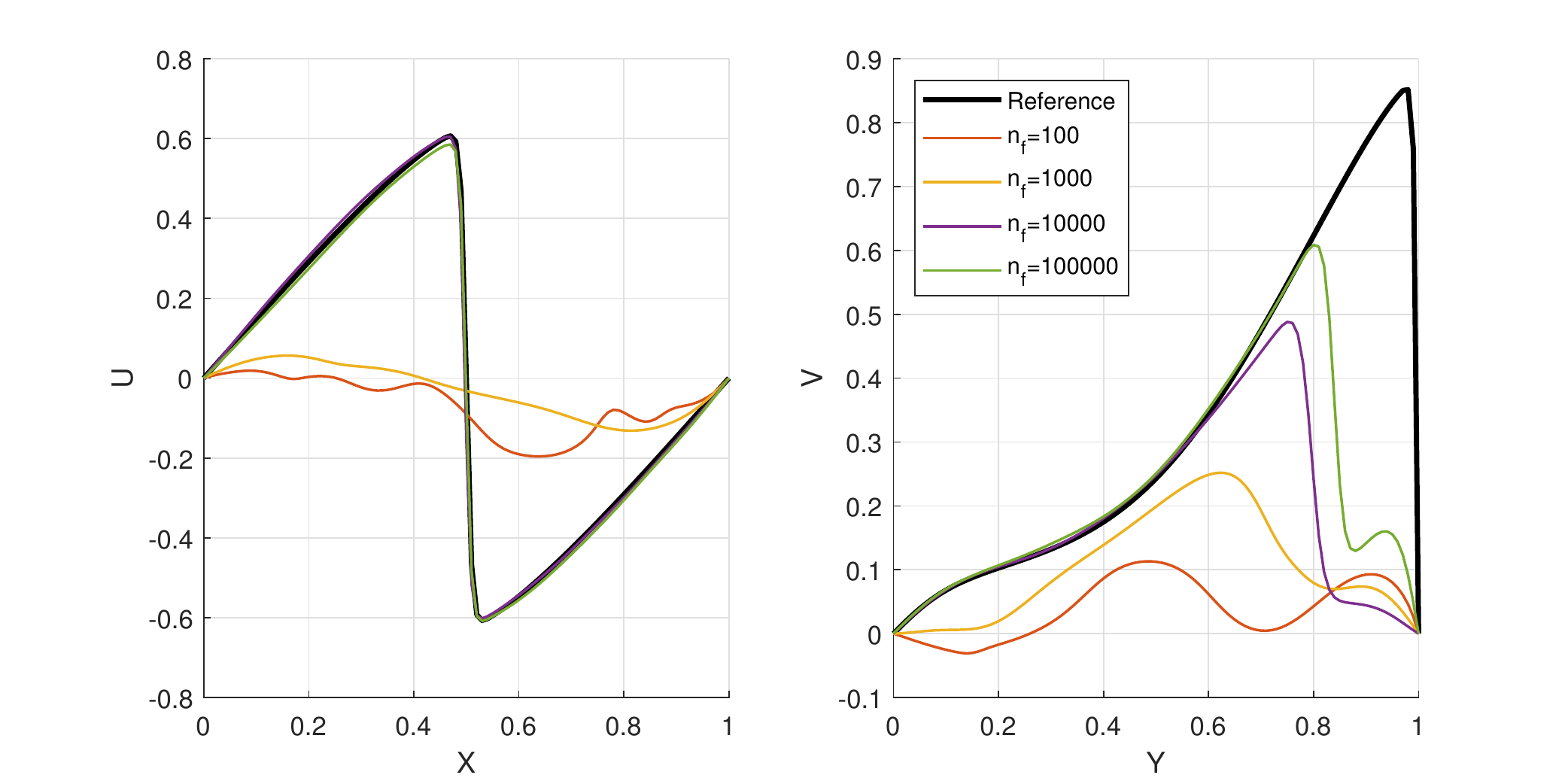}
	\caption{(Left) Plot of $U$ with respect to $x$ at $y=0.25$ and $t=0.5$ for $n_s=0$ and various values of $n_f$. (Right) Plot of $V$ with respect to $y$ at $x=0.25$ and $t=0.5$ for $n_s=0$ and various values of $n_f$.}
	\label{fig:ns0_compare}
\end{figure}

\begin{figure}[!ht]
	\centering
	\includegraphics[width=1\textwidth]{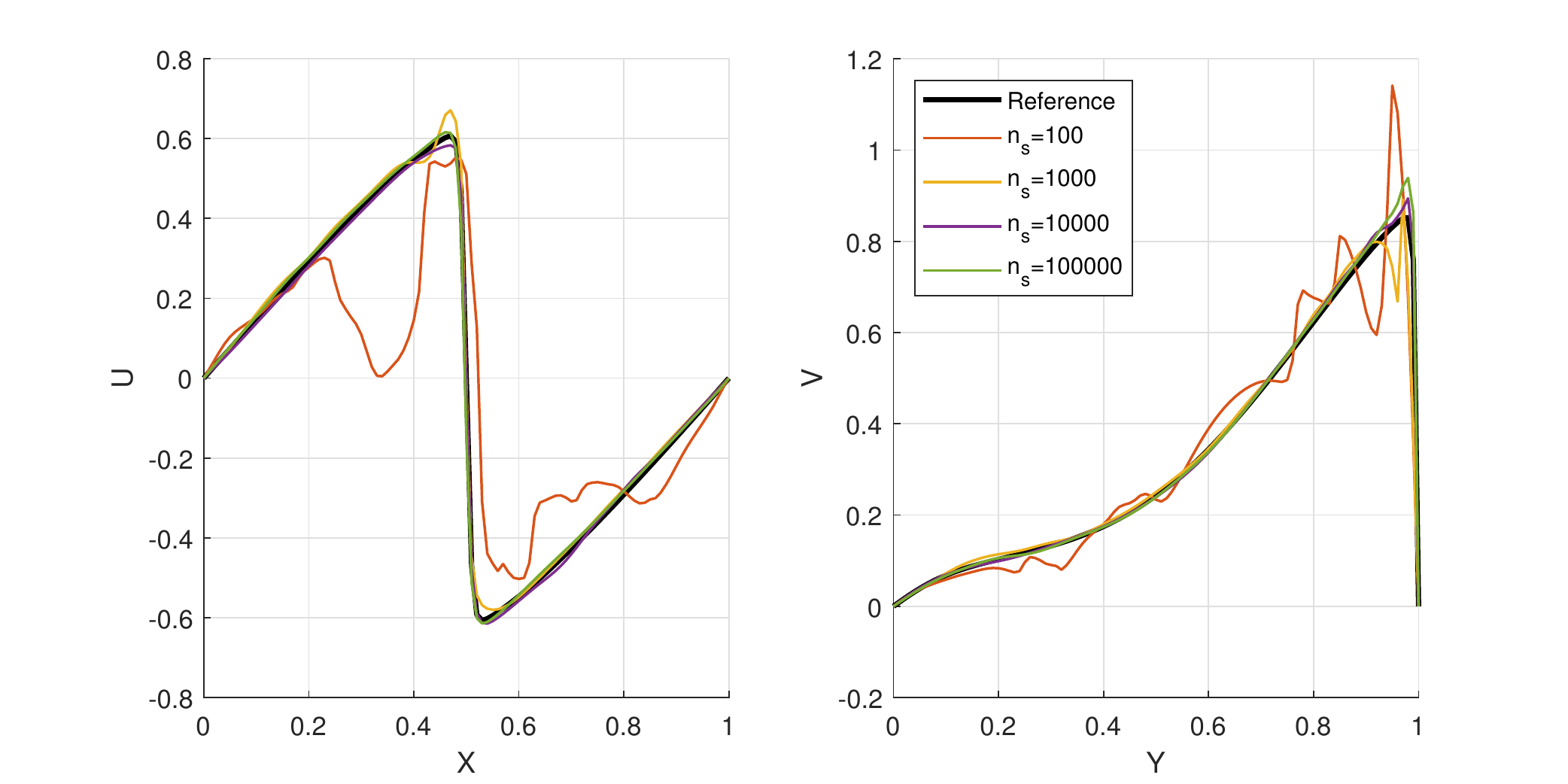}
	\caption{(Left) Plot of $U$ with respect to $x$ at $y=0.25$ and $t=0.5$ for $n_f=0$ and various values of $n_s$. (Right) Plot of $V$ with respect to $y$ at $x=0.25$ and $t=0.5$ for $n_f=0$ and various values of $n_s$.}
	\label{fig:nf0_compare}
\end{figure}

For completeness, a similar behaviour can be seen when the same plots are made for both $n_s=0$ and varying $n_f$ and for $n_f=0$ and varying $n_s$, shown in Figures~\ref{fig:ns0_compare}~and~\ref{fig:nf0_compare}, respectively. For the cases with no sample data ($n_s=0$), as the number of function evaluation points increases, the values of $U$ and $V$ move away from null values and approximate the reference solution. On the other hand, when the model has no physical reference ($n_f=0$), the predictions quickly approximate the reference data, even for low values of $n_s$; however, only models trained with larger values of $n_s$ can reach a smooth prediction, which more closely observes the governing equations.

\begin{figure}[!ht]
	\centering
	\includegraphics[width=1\textwidth]{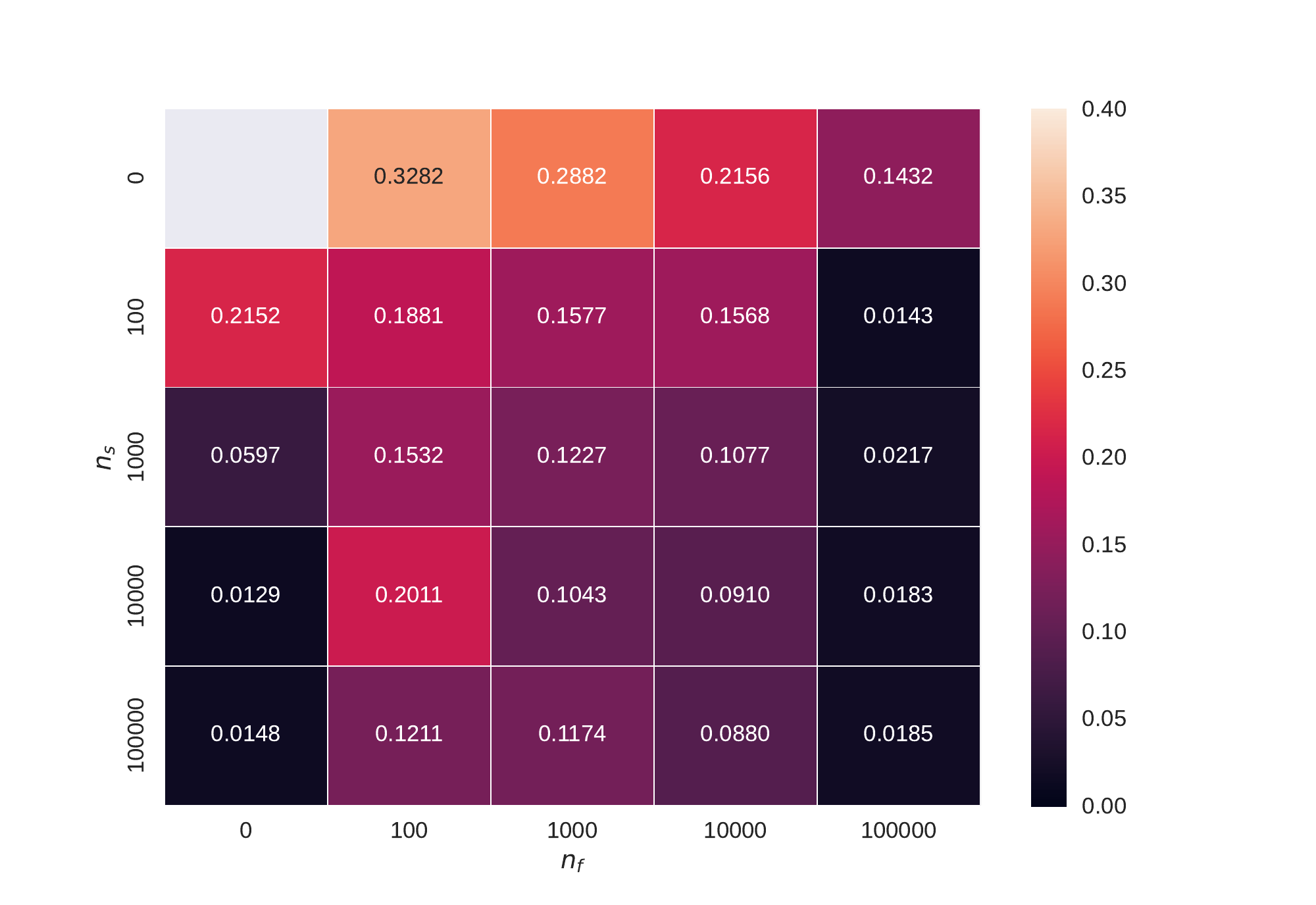}
	\caption{Root mean square of the error for different combinations of function evaluation points ($n_f$) and solution sample points ($n_s$).}
	\label{fig:heat_err}
\end{figure}

The errors of each model concerning the reference data were estimated by computing the model at a grid of uniformly spaced points and comparing it to the reference. Figure~\ref{fig:heat_err} summarizes the root mean square of the errors of each case. It is possible to note that the physics-less case often produces lower errors when compared to the physics-informed model, especially at the larger values of $n_s$. Nonetheless, by further increasing $n_f$, the errors decrease considerably. This phenomenon can be explained by looking back at Figure~\ref{fig:nf_compare}, where we have argued that, despite reaching values that are closer to the reference data in an absolute sense, the solution produced by the physics-less model is not smooth, which incurs a larger non-observance of the governing equations.

\subsection{Computational Cost}

\begin{figure}[!ht]
	\centering
	\includegraphics[width=1\textwidth]{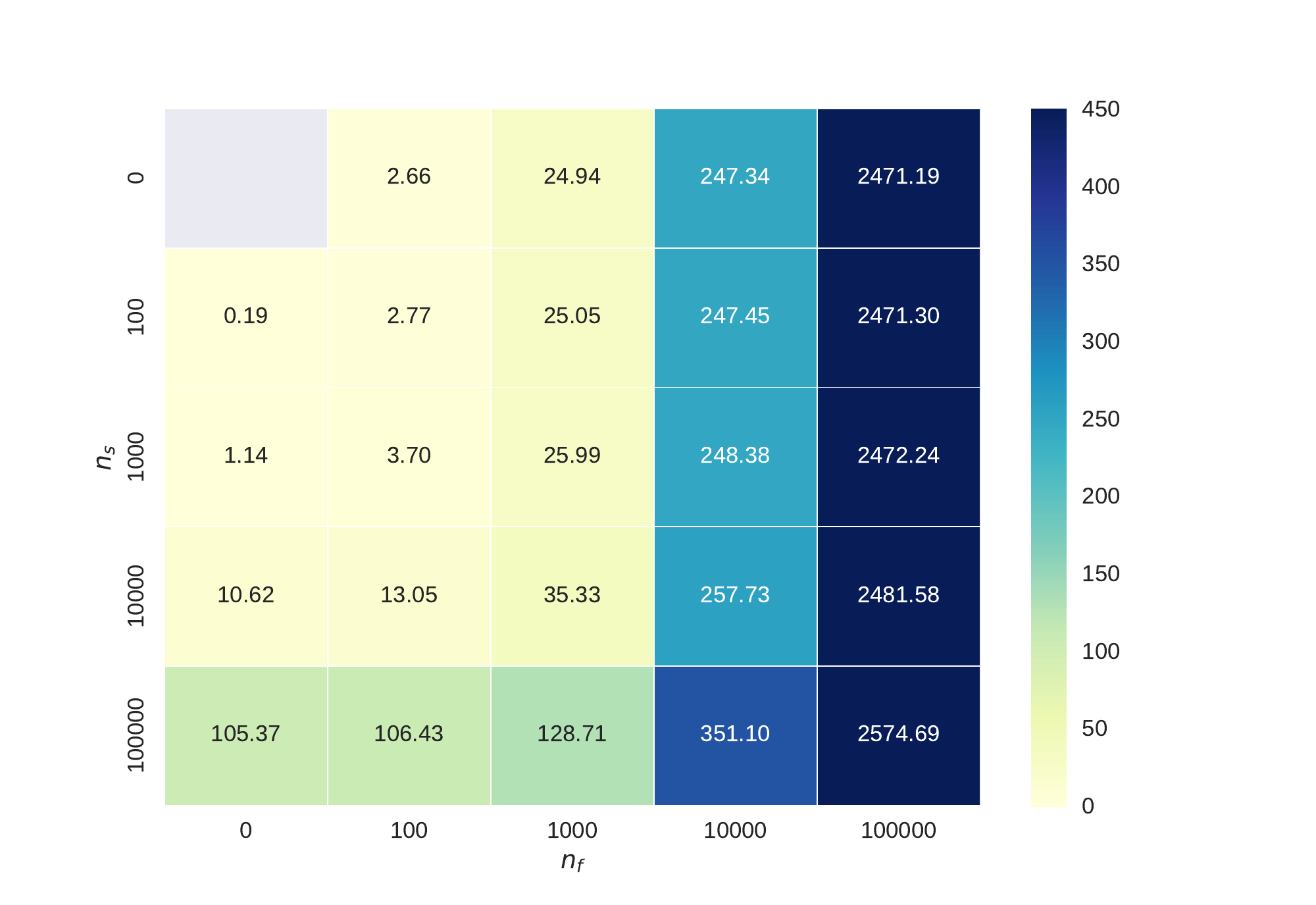}
	\caption{Maximum allocated GPU memory,in megabytes, for different combinations of function evaluation points ($n_f$) and solution sample points ($n_s$).}
	\label{fig:heat_memory}
\end{figure}

Each case's memory footprint was measured using PyTorch's native method \verb|cuda.max_memory_allocated|, which returns the largest amount of GPU memory allocated during execution. Naturally, increasing either $n_f$ or $n_s$ has caused more memory to be allocated. Both values presented a roughly linear relation to the allocation size. Nonetheless, the number of function evaluations had a much larger impact than the number of sample points. This result is caused by the much larger complexity in computing the residuals of the governing equations as opposed to simply comparing the model's output to the reference data. The case with $n_f=0$ and $n_s=100000$ has allocated just over 100 MB of memory, while the $n_f=100000$ and $n_s=0$ has needed over 2.5 GB. Figure~\ref{fig:heat_memory} shows the values for each combination of $n_f$ and 

\section{Discussion}
\label{sec:discussion}

In this paper, we implemented a Physics-Informed Neural Network to solve the 2D Burgers equation. During training, this model can use both function evaluation points, where the residual of the governing equations is sought to be minimized, and solution sample points, where the solution's value is known and training seeks to minimize the difference between the model's output and the reference data. The network was trained with different amounts of function evaluation points ($n_f$) and solution sample points ($n_s$), and the resulting models were compared.

By looking at the residual of the governing equations, we improved the models by increasing either $n_f$ or $n_s$. Nonetheless, we have observed that this improvement happens in different manners for each variable. Increasing $n_s$ causes the model to approximate the reference values quickly. However, only larger sample sizes can reach smooth curves, which more closely follow the governing equations. On the other hand, by setting $n_s$ to zero and relying only on function evaluation points, the solution obtained is usually smooth, but it can only approach the actual solution for larger batch sizes ($n_f \ge 10000$ in our case).

These results can be beneficial for scenarios where the availability of reference data is limited. Adding function evaluation points allows one to improve the model without requiring additional data. Nonetheless, care must be taken so that enough function evaluation points are used. Otherwise, training may lead to solutions that do not follow the reference data, despite having lower residuals in the governing equations.

For future works, we suggest investigating how the balance between the physics-based loss and the data-based loss influences the resulting model. Perhaps different loss functions would be able to take advantage of the lowered governing equation residuals offered by adding function evaluation points without the downside of causing the model to drift away from the reference data, as we have observed at low values of $n_f$. 

\section*{Acknowledgements}

This work was carried out at the Center for Artificial Intelligence (C4AI-USP), with support by the São Paulo Research Foundation (FAPESP) under grant number 2019/07665-4, and by the IBM Corporation. This work is also supported in part by FAPESP under grant number 2020/16746-5, the Brazilian National Council for Scientific and Technological Development (CNPq) under grant numbers 310085/2020-9, 310127/2020-3, 312180/2018-7, Coordination for the Improvement of Higher Education Personnel (CAPES, Finance Code 001), and by \textit{Ita\'{u} Unibanco S.A.} through the \textit{Programa de Bolsas Ita\'{u}} (PBI) program of the \textit{Centro de Ci\^{e}ncia de Dados} (C$^2$D) of \textit{Escola Polit\'{e}cnica} of USP.

\section*{Preprint disclaimer}

This preprint has not undergone peer review or any post-submission improvements or corrections. The Version of Record of this contribution is published in the Lecture Notes in Computer Science book series (LNAI,volume 13654), and is available online at \href{https://doi.org/10.1007/978-3-031-21689-3_28}{https://doi.org/10.1007/978-3-031-21689-3\_28}

%
%
%
\bibliographystyle{splncs04}
\bibliography{references}

\begin{thebibliography}{10}
\providecommand{\url}[1]{\texttt{#1}}
\providecommand{\urlprefix}{URL }
\providecommand{\doi}[1]{https://doi.org/#1}

\bibitem{beucler2021}
Beucler, T., Pritchard, M., Rasp, S., Ott, J., Baldi, P., Gentine, P.:
  Enforcing {{Analytic Constraints}} in {{Neural Networks Emulating Physical
  Systems}}. Physical Review Letters  \textbf{126}(9) (2021).
  \doi{10.1103/PhysRevLett.126.098302}

\bibitem{fukami2020}
Fukami, K., Fukagata, K., Taira, K.: Machine-learning-based spatio-temporal
  super resolution reconstruction of turbulent flows. Journal of Fluid
  Mechanics  \textbf{909} (2020). \doi{10.1017/jfm.2020.948}

\bibitem{he2016}
He, K., Zhang, X., Ren, S., Sun, J.: Deep {{Residual Learning}} for {{Image
  Recognition}}. In: 2016 {{IEEE Conference}} on {{Computer Vision}} and
  {{Pattern Recognition}} ({{CVPR}}). pp. 770--778 (Jun 2016).
  \doi{10.1109/CVPR.2016.90}

\bibitem{karniadakis2021}
Karniadakis, G.E., Kevrekidis, I.G., Lu, L., Perdikaris, P., Wang, S., Yang,
  L.: Physics-informed machine learning. Nature Reviews Physics  \textbf{3}(6),
   422--440 (2021). \doi{10.1038/s42254-021-00314-5}

\bibitem{kashinath2021}
Kashinath, K., Mustafa, M., Albert, A., Wu, J.L., Jiang, C., Esmaeilzadeh, S.,
  Azizzadenesheli, K., Wang, R., Chattopadhyay, A., Singh, A., Manepalli, A.,
  Chirila, D., Yu, R., Walters, R., White, B., Xiao, H., Tchelepi, H.A.,
  Marcus, P., Anandkumar, A., Hassanzadeh, P., {Prabhat}: Physics-informed
  machine learning: {{Case}} studies for weather and climate modelling.
  Philosophical Transactions of the Royal Society A: Mathematical, Physical and
  Engineering Sciences  \textbf{379}(2194) (2021). \doi{10.1098/rsta.2020.0093}

\bibitem{lele1992}
Lele, S.K.: Compact finite difference schemes with spectral-like resolution.
  Journal of Computational Physics  \textbf{103}(1),  16--42 (Nov 1992).
  \doi{10.1016/0021-9991(92)90324-R}

\bibitem{li2021}
Li, Z., Kovachki, N.B., Azizzadenesheli, K., Liu, B., Bhattacharya, K., Stuart,
  A., Anandkumar, A.: Fourier {{Neural Operator}} for {{Parametric Partial
  Differential Equations}}. In: International {{Conference}} on {{Learning
  Representations}} (2021)

\bibitem{nair2019}
Nair, A.G., Yeh, C.A., Kaiser, E., Noack, B.R., Brunton, S.L., Taira, K.:
  Cluster-based feedback control of turbulent post-stall separated flows.
  Journal of Fluid Mechanics  \textbf{875}(M),  345--375 (2019).
  \doi{10.1017/jfm.2019.469}

\bibitem{raissi2019}
Raissi, M., Perdikaris, P., Karniadakis, G.E.: Physics-informed neural
  networks: {{A}} deep learning framework for solving forward and inverse
  problems involving nonlinear partial differential equations. Journal of
  Computational Physics  \textbf{378},  686--707 (2019).
  \doi{10.1016/j.jcp.2018.10.045}

\bibitem{read2019}
Read, J.S., Jia, X., Willard, J., Appling, A.P., Zwart, J.A., Oliver, S.K.,
  Karpatne, A., Hansen, G.J., Hanson, P.C., Watkins, W., Steinbach, M., Kumar,
  V.: Process-{{Guided Deep Learning Predictions}} of {{Lake Water
  Temperature}}. Water Resources Research  \textbf{55}(11),  9173--9190 (2019).
  \doi{10.1029/2019WR024922}

\bibitem{sun2020}
Sun, L., Gao, H., Pan, S., Wang, J.X.: Surrogate modeling for fluid flows based
  on physics-constrained deep learning without simulation data. Computer
  Methods in Applied Mechanics and Engineering  \textbf{361},  112732 (Apr
  2020). \doi{10.1016/j.cma.2019.112732}

\bibitem{willard2020}
Willard, J., Jia, X., Xu, S., Steinbach, M., Kumar, V.: Integrating
  {{Scientific Knowledge}} with {{Machine Learning}} for {{Engineering}} and
  {{Environmental Systems}}  \textbf{1}(1),  1--34 (Mar 2020)

\bibitem{wu2020}
Wu, M., Stefanakos, C., Gao, Z.: Multi-step-ahead forecasting of wave
  conditions based on a physics-based machine learning ({{PBML}}) model for
  marine operations. Journal of Marine Science and Engineering  \textbf{8}(12),
   1--24 (2020). \doi{10.3390/jmse8120992}

\bibitem{xu2015}
Xu, T., Valocchi, A.J.: Data-driven methods to improve baseflow prediction of a
  regional groundwater model. Computers and Geosciences  \textbf{85},  124--136
  (2015). \doi{10.1016/j.cageo.2015.05.016}

\end{thebibliography}
\end{document}